\documentclass[twocolumn,showpacs,aps,pre,
groupedaddress,amssymb,amsmath,nobalancelastpage]{revtex4}

\usepackage{graphicx}
\usepackage{longtable}

\begin{document}


\title{Dependence of Striped Domain Growth on Background Wavenumber}

\author{Carina Kamaga}
\author{Denis Funfschilling}
\author{Michael Dennin}
\affiliation{Department of Physics and Astronomy, University of
California at Irvine, Irvine, California 92697-4575}

\date{\today}

\begin{abstract}

We report on the growth of domains of standing waves in
electroconvection in a nematic liquid crystal. An ac voltage is
applied to the system, forming an initial state that consists of
travelling striped patterns with two different wavenumbers, zig
and zag rolls. The standing waves are generated by suddenly
applying a periodic modulation of the amplitude of the applied
voltage that is approximately resonant with the travelling
frequency of the pattern. By varying the modulation frequency, we
are able to vary the steady-state, average wavenumber. We
characterize the growth by measuring the structure factor for the
pattern, the amplitude of the pattern, and the domain-wall length.
An interesting feature of the growth is that the evolution of the
domain-wall length is strongly affected by changing the selected
wavenumber; whereas, the evolution of the structure factor is less
sensitive to such changes.

\end{abstract}

\pacs{47.54.+r,64.60.Cn}

\maketitle

\section{Introduction}

When a system experiences a sufficiently large, sudden change of
an external parameter, referred to as a quench, the current state
of the system will lose stability and a transition will occur to a
new state. When there is more than one possibility for the new
state (such as spin up or down in a magnetic system), domains will
form. At late times, these domains coarsen, or grow. The dynamics
of this coarsening process has been the subject of substantial
research and is important in many applications \cite{B94}. In
general, the domains are characterized by a single length scale,
$l(t)$, that grows according to a power-law, $l(t) \sim t^n$. Here
$n$ is referred to as the {\it growth exponent}. This process is
usually referred to as {\it domain coarsening}. Some examples of
systems for which domain coarsening has been studied in detail
\cite{B94}, either theoretically or experimentally, include the
Ising model, binary fluids, liquid crystals, diblock copolymers
\cite{HAC00,HCSH02}, and electroconvection \cite{PD01}.

There are a number of ways to categorize systems undergoing
coarsening. First, one must distinguish between cases where the
order parameter is conserved, such as in a binary fluid, and where
it is not conserved, such as in the Ising model. Second, the
domains that are growing may be uniform, such as in both the Ising
model and binary fluids, or they may consist of a pattern, such as
stripes in the case of diblock copolymers and electroconvection.
Finally, the initial and final states can either be thermodynamic
equilibrium states or steady states of the system. Most of the
studies have focused on the case where the initial state prior to
the quench and the final state well after the quench are
thermodynamic equilibrium states \cite{B94}. We will refer to this
type of system as a {\it thermodynamic system}. For isotropic,
thermodynamic systems, where the domains are uniform, the
exponents are well known. For systems with a non-conserved order
parameter, $n = 1/2$. Systems with a conserved order parameter
have $n = 1/3$. In this report, we focus on coarsening in a system
in which the domains are composed of stripes. An added feature of
the system we consider is that the initial and final states are
{\it not} states of thermodynamic equilibrium, though they are
steady-states of the system. We refer to this as a driven system.

For systems in which the domains are composed of stripes,
theoretical studies of the power-law scaling of the late-time
domain growth has focused on simulations and analysis of the
Swift-Hohenberg equation
\cite{EVG92,CM95,CB98,BV01,BV02,TNS02,QM02}. Experiments have been
performed in both a thermodynamic system, diblock copolymers
\cite{HAC00,HCSH02}, and a driven system, electroconvection
\cite{PD01,KID03}. This work has raised two related issues. First,
two length scales are important in the problem: the wavelength of
the pattern and the scale of a domain. This has an interesting
impact on the measured growth exponents. Traditional studies of
domain growth focus on the structure factor, $S(k)$. This measures
both variations in wavenumber and domain size, with information
about both lengths scales mixing in a nontrivial way. For striped
systems, one can also consider orientational correlation
functions, defect densities, or domain wall lengths. These
quantities are independent of the local wavenumber and are
expected to provide information about domain size only.

Numerical simulations have considered both the scaling of the
structure factor, $S(k)$, of the system and the orientational
correlation function. The evolution of the structure factor is
found to be consistent with a length scale that grows as $t^{1/5}$
for both potential and nonpotential versions of the
Swift-Hohenberg equation \cite{CM95,CB98}. However, the growth
exponent determined from the orientational correlation function is
$1/4$ for the potential case and $1/2$ for the nonpotential case
\cite{CM95}. The orientational correlation function and defect
densities are measured experimentally in diblock copolymers
\cite{HAC00}. The results agree with the potential case of the
Swift-Hohenberg model, $n = 1/4$. For a particular case of
electroconvection, it was shown that the wavenumber and domains
evolve differently \cite{PD01}. It was found that the growth
exponent measured using $S(k)$ is consistent with $n = 1/5$ and
measurement of the domain size gave $n = 1/4$. This is in
agreement with simulations for the potential case \cite{CM95}. In
addition, the distribution of local wavenumbers was measured
separately. This distribution was essentially independent of time,
providing an explanation for the difference between $S(k)$ and
domain wall measurements.

The second issue associated with striped domains is the
interaction between defects in the pattern and the background
stripe pattern. This has been studied using the Swift-Hohenberg
equation. One important consequence is that the growth of domains
can be strongly affected by pinning of grain boundaries
\cite{BV01,BV02}. In these simulations, the measured growth
exponents depends on the depth of the quench, something that does
not occur in uniform systems. The growth exponents are also found
to depend on the level of noise in the system \cite{EVG92,TNS02}.
This is attributed to the fact that noise can provide a source of
``depinning'' for the defects. Finally, the evolution of the
average wavenumber provides a possible explanation for differences
in simulations of the potential and nonpotential versions of the
Swift-Hohenberg equations. The average wavenumber in each case
evolves to different values, and this was suggested as one
possible explanation for the two different growth exponents
\cite{CM95}.

We report on an experimental study of the impact of changes in the
average wavenumber on the growth of domains. We utilize a
particular example of electroconvection that is ideally suited for
controlling the average wavenumber of the pattern:
electroconvection in the nematic liquid crystal I52
\cite{DCA98,D00b}. A nematic liquid crystal is a fluid in which
the molecules are aligned on average \cite{GP93}. The axis along
which the average alignment points is referred to as the director.
For electroconvection \cite{KP95}, a nematic liquid crystal is
placed between two glass plates that have been treated so that the
director is everywhere parallel to the plates. An ac voltage is
applied perpendicular to the plates. Above a critical value of the
voltage, a pattern of stripes forms. Because the system is
anisotropic (the director selects a preferred axis), the states
are usually characterized by the angle between the wavevector of
the pattern and the undistorted director, $\theta$. States with
the same wavenumber and angle $\pm \theta$ are degenerate and
referred to as zig and zag, respectively. For the system discussed
in this paper, the initial pattern consists of four travelling
stripe states \cite{DTKAC96}: right- and left-travelling zig and
zag rolls. It has been established for this system that above a
critical value for a modulation of the applied voltage at twice
the travelling frequency, the system consists of either standing
zig or zag rolls \cite{D00b}. Therefore, a sudden quench of the
system from below this critical value to above it results in the
nucleation of domains of standing zig and zag rolls, which proceed
to coarsen.

Previous measurements of growth exponents for this system focused
on quenches that used exactly twice the travelling frequency.
Reference~\cite{PD01} reports a value of $n = 1/4$ for the growth
of the domain size and $n = 1/5$ for the scaling of $S(k)$. A
single initial experiment for a modulation frequency slightly
different from twice the travelling frequency suggested similar
behavior. However, a systematic exploration of changes in the
wavenumber, by varying the modulation frequency, is possible for
this system \cite{D00b}, and has not been carried out. In this
paper, we report on such a study. The rest of the paper is
organized as follows. Section II provides the details of the
experimental system and the techniques used to analyze the domain
growth. Section III presents the results of the experiment.
Section IV provides a discussion of the results.

\section{Experimental Details}

We used the liquid crystal I52 \cite{FGWP89} doped with iodine.
Due to the nature of I52, $8\%$ iodine by weight was used
initially. However, a significant fraction of the iodine
evaporates before the cells are filled, so the final concentration
is not well controlled. We used commercial liquid crystal cells
\cite{EHCO} composed of two transparent glass plates coated with a
transparent conductor, indium-tin oxide. The cells were 2.54 cm x
2.54 cm, with the conductor forming a square 1 cm x 1 cm in the
center of the cell. The glass was also treated with a rubbed
polymer to obtain uniform planar alignment of the director. We
define the direction parallel to the undistorted director to be
the x axis and the direction perpendicular to the undistorted
director to be the y axis.

The apparatus is described in detail in Ref. \cite{D00a}. It
consists of an aluminum block with glass windows. The temperature
of the block was kept constant to $\pm 0.001\ ^{\circ}{\rm C}$.
Polarized light is applied from the bottom of the cell. Above the
cell, there is a ccd camera to capture the image of patterns using
the standard shadowgraph technique \cite{RHWR89}.

We apply an ac voltage of the form $V(t) = \sqrt{2}[V_o + V_m
cos(\omega_m t)]cos(\omega_d t)$ perpendicular to the glass
plates. Here $V_o$ is the amplitude of the applied voltage in the
absence of modulation, $V_m$ is the modulation amplitude,
$\omega_m$ is the modulation frequency, and $\omega_d$ is the
driving frequency. For all of the experiments reported on here,
$\omega_d/2\pi = 25\ {\rm Hz}$. The critical voltage, $V_c$ is
defined as the value of $V_o$ for which the system makes a
transition from a uniform state to the superposition of travelling
rolls {\it in the absence of modulation}. There are two relevant
control parameters: $\epsilon = (V_o/V_c)^2 - 1$ and $b =
V_m/V_c$.

\begin{figure}
\includegraphics[width=3.0in]{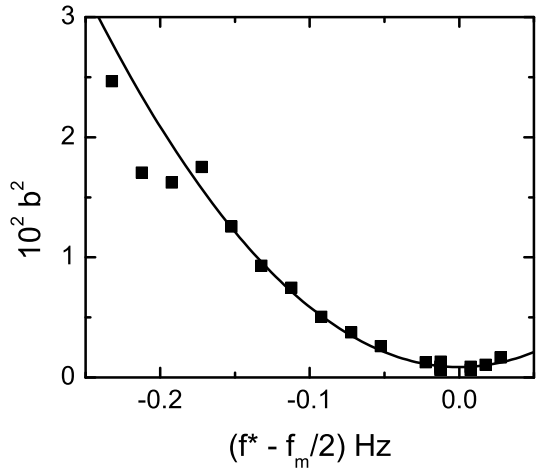}
\caption{\label{onset} Onset curve for standing waves at a value
of $\epsilon = 0.04$ as a function of $\delta f = f^* - f_m/2$.
Here $f^*$ is the natural travel frequency of the pattern at
$\epsilon = 0.04$ and $f_m$ is the modulation frequency. The
parameter $b = V_m/V_c$, where $V_m$ is the modulation voltage and
$V_c$ is the critical voltage for the onset of a pattern in the
absence of any modulation. The solid line is a fit to the data
used to determine $f^*$.}
\end{figure}

For the experiments discussed here, the system is initially at
$\epsilon = 0.04$ and $b = 0$, i.e. above onset with no
modulation. As mentioned in Sec. I, there are only four modes
needed to describe the observed convection rolls: zig and zag,
right and left travelling oblique rolls.  This state is an example
of spatiotemporal chaos in that the amplitude of the four modes
varies irregularly in space and time \cite{DCA96}. For each
$\epsilon$ and $\omega_m$, there exists a critical value of $b$,
$b_c$, above which the system consists of standing zig and zag
rolls. A quench corresponds to a sudden change of $b$ to some
value $b > b_c$. At and below $V_c$, the modulation frequency with
the strongest coupling to the pattern, and correspondingly
smallest value of $b_c$, is twice the Hopf frequency, $\omega_h$.
The Hopf frequency is the travelling wave frequency at $V_c$. As
one increases $\epsilon$, the natural frequency of the pattern
shifts. Because we worked at a finite value of $\epsilon$, the
smallest value of $b_c$ occurs for a modulation frequency,
$\omega_m = 2\omega^*$, where $\omega^*$ is the natural frequency
of the pattern at the given value of $\epsilon$. Because it is the
relative shift in frequency that matters, we will consider the
behavior as a function of $\delta f = f^* - f_m/2$, where $f^* =
\omega^*/2\pi$ and $f_m = \omega_m/2\pi$. Figure~\ref{onset} is a
plot of $b_c$ as a function of $\delta f$ for $\epsilon = 0.04$.
We focus almost exclusively on the negative values of $\delta f$
because for $\delta f > +0.01\ {\rm Hz}$, the pattern stabilized
by modulation is standing waves that are a superposition of zig
and zag rolls, and not individual domains of standing zig and zag
rolls.

Quenches were performed for a range of $\delta f$. The final value
of $b$ in each case was selected so that the quench depths all
corresponded to $\Delta b = 0.02$, as measured from the onset of
standing waves (solid curve in Fig.~\ref{onset}). The camera was
triggered to take the images at the same point in the modulation
cycle. This ensured that the pattern was imaged at the same point
in the standing wave cycle. Each image covered a region of $4.1\
{\rm mm} \times 3.1\ {\rm mm}$ of the cell. Time is scaled by the
director relaxation time, $\tau_d \equiv \gamma_1 d^2/(\pi^2
K_{11}) = 0.2\ {\rm s}$. Here $\gamma_1$ is a rotational
viscosity; $K_{11}$ is the splay elastic constant; and $d$ is the
thickness of the cell.

\begin{figure}
\includegraphics[width=3.0in]{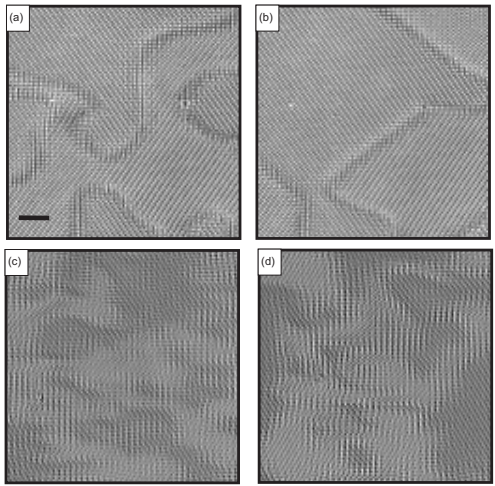}
\caption{\label{images} Four images of domain growth. Images (a)
and (b) are with a modulation frequency $\delta f = 0.008\ {\rm
Hz}$, at $t = 96\ {\rm s}$ and $t = 600\ {\rm s}$ after the
quench, respectively. Images (c) and (d) are with a modulation
frequency $\delta f = -0.23\ {\rm Hz}$, at $t = 107\ {\rm s}$ and
$t = 667\ {\rm s}$ after the quench, respectively. The solid bar
in (a) is 0.2 mm.}
\end{figure}

An example of a small region of the pattern after a quench is
given in Fig.~\ref{images}. Two different quenches are
illustrated: $\delta f = 0.008\ {\rm Hz}$ and $\delta f = -0.23\
{\rm Hz}$. The images for $\delta f = 0.008\ {\rm Hz}$ illustrate
the type of patterns on which we focused: standing zig and zag
rolls separated by domain walls. The domain walls are relatively
``thick'', and the superposition of zig and zag rolls in a domain
wall is visible. The quench at $\delta f = -0.23\ {\rm Hz}$
illustrates qualitatively different behavior. In this case, there
are regions of superposition that are ``domains'' in their own
right, though most of the regions of superposition can be
considered to be domain walls. As we will show, this qualitative
difference is apparent in various characterizations of the growth.
As the focus of the paper is the impact of varying the wavenumber
on the domain growth in the zig/zag system, we will only report on
this single quench in the regime where domains of the
superposition of zig and zag rolls occurs. Further increasing
$|\delta f|$ results in patterns that are entirely the
superposition of zig and zag rolls. Future work will consider in
more detail this other class of patterns.

A number of features of the domains were analyzed. First, the
power spectrum, $S(k)$, of each image was computed. This was used
to determine the average wavenumber ($<k> \equiv \int k S(k)
dk/\int S(k)dk$) for the zig and zag rolls. The widths of the
fundamental peak in $S(k)$ in both the x ($\sigma_x$) and y
($\sigma_y$) directions were determined using the second moment of
$S(k)$. This gives $\sigma_x \equiv \sqrt{<k_x^2> - <k_x>^2}$,
with $<k_x^2> \equiv \int k_x^2 S(k) dk/\int S(k)dk$, and a
similar definition for $\sigma_y$. (Note: all of the integrals
were taken as sums over a finite region around the fundamental
peak, as described in Ref.~\cite{PD01}.) $\sigma_x$ and $\sigma_y$
contain information about the typical size of a domain and the
distribution of wavenumbers, in the respective directions.

\begin{figure}
\includegraphics[width=3.0in]{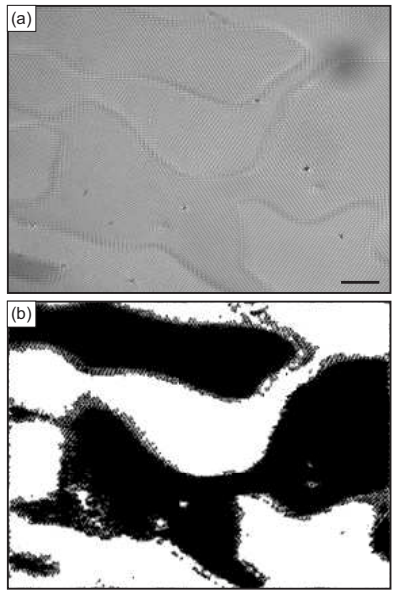}
\caption{\label{anglemap} (a) Image taken $384\ {\rm s}$ after a
quench with $\delta f = 0.008\ {\rm Hz}$. (b) The same image
converted to an ``angle'' map. The black and white regions are zig
and zag rolls, respectively. The gray regions are the boundaries
between domains. The scale bar in (a) represents $0.4\ {\rm mm}$.}
\end{figure}

We also analyzed the images by first converting them to a scaled
image of zig and zag domains, referred to as the {\it angle map}.
This was done using the procedure detailed in Ref.~\cite{PD01}.
Briefly, the local angle of the wavenumber is computed using the
algorithm reported in Ref.~\cite{EMB98}. This is used to produce
an image with black and white domains representing regions of zig
and zag, respectively. Regions of superposition of zig and zag
rolls (the domains wall for most of the quenches) are represented
by gray. Figure~\ref{anglemap} illustrates a typical image
(Fig.~\ref{anglemap}a) and the results after conversion to an
angle map (Fig.~\ref{anglemap}b). Once these images are produced,
two measures of the domain growth that are independent of
variations in the local wavenumber are computed. First, one can
determine the total domain wall length per viewing area, $L$. This
quantity scales as the inverse of a characteristic domain size
(the area of the viewing scales as a length squared). In our case,
$L$ was defined as the total number of gray pixels divided by the
total area of the image. For all of the quenches except the ones
at $\delta f = -0.23\ {\rm Hz}$, the width of the domain walls was
constant during the domain growth. Therefore, this is a consistent
measure of the total domain wall length. For the quench at $\delta
f = -0.23\ {\rm Hz}$, we still used this definition for
consistency, even though it represents a transitional behavior. In
this case, some of the regions of superposition of zig and zag
rolls (gray pixels) are more appropriately considered domains in
their own right. Still, most of the gray pixels corresponded to
domain walls between zig and zag domains. We also computed the
power spectrum of the angle map image, $S(q)$. In this case, the
peak is centered at $q = 0$, as the angle map domains are uniform.
Therefore, the width of the peak is related to the typical domain
size only. Again, we separately considered the width in the x
direction ($w_x$) and y direction ($w_y$), where the widths are
defined in the same manner as for $S(k)$.

\section{Results}

\begin{figure}
\includegraphics[width=3.0in]{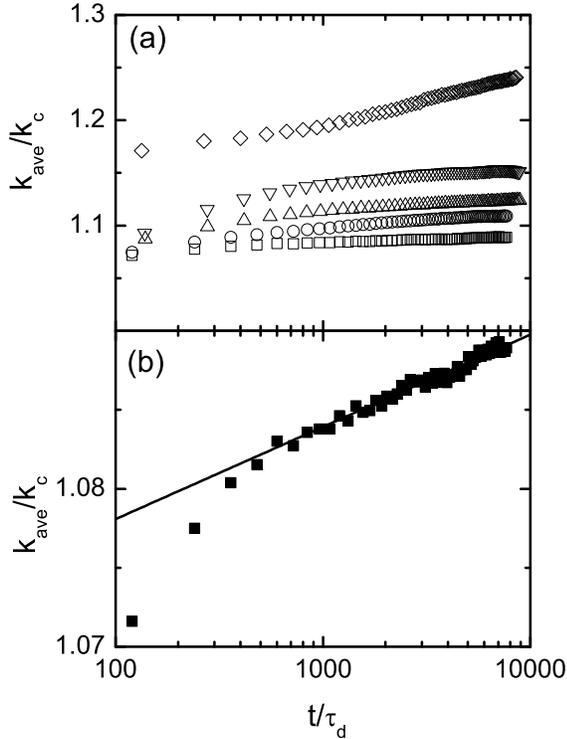}
\caption{\label{wavenumber} (a) Average wavenumber for the zig
standing rolls as a function of time after the quench. Time is
scaled by the director relaxation time, $\tau_d$. The symbols are
as follows: $(\square)\ \delta f = 0.008\ {\rm Hz}$, $(\circ)\
\delta f = -0.012\ {\rm Hz}$, $(\vartriangle)\ \delta f = -0.022\
{\rm Hz}$, $(\triangledown)\ \delta f = -0.072\ {\rm Hz}$, and
$(\lozenge)\ \delta f = -0.23\ {\rm Hz}$. The wavenumber for the
zag rolls shows similar behavior. (b) An expanded view of the
average wavenumber for $\delta f = 0.008\ {\rm Hz}$. The solid
line is a guide to the eye illustrating the regime where the
evolution of $<k>$ is consistent with logarithmic growth in time.}
\end{figure}

Figure~\ref{wavenumber} shows the time evolution of the average
wavenumber ($<k>$) for the different quenches of interest. A
number of features of the evolution are evident. First, as
claimed, the steady-state value of the $<k>$ is dependent on
$\delta f$ (see Fig.~\ref{wavenumber}a). However, for all
quenches, the $<k>$ is still evolving even at late times. After
$t/\tau_d \approx 1000$, the evolution is consistent with a very
slow logarithmic growth: $<k> = a \log_{10}(t/\tau_d)$, with $a
\approx 0.01$. This is illustrated in Fig.~\ref{wavenumber}b.
However, for all but the largest value of $|\delta f|$, there is a
weak dependence on when the logarithmic growth begins (with values
ranging from $2.7 < \log_{10}(t/\tau_d) < 3.1$) and on the
coefficient ($a$) of the logarithmic growth (with values ranging
from $0.006$ to $0.02$. In general, for larger $|\delta f|$,
evolution consistent with logarithmic time dependence occurs at a
later time. As referred to in Sec.~II, for $\delta f = -0.23\ {\rm
Hz}$, the pattern is qualitatively different. In this case, the
evolution of $<k>$ is also very different, with logarithmic growth
setting in significantly later ($\log_{10}(t/\tau_d) \approx 3.5$)
and with a coefficient of $0.05$.

\begin{figure}
\includegraphics[width=3.0in]{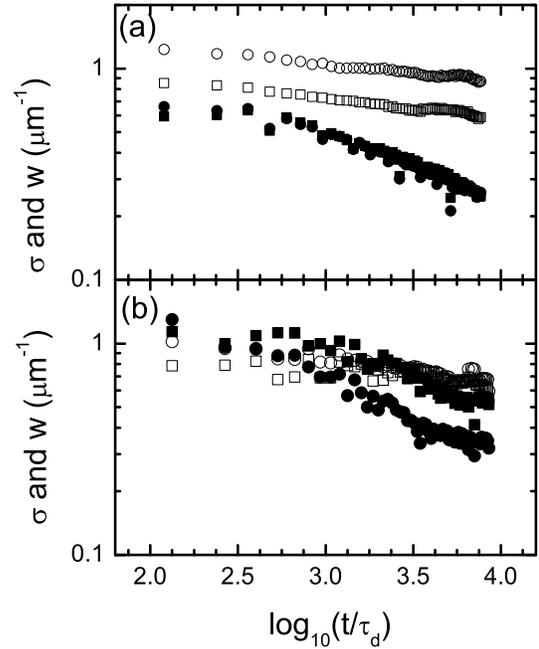}
\caption{\label{sigma} Width of the power spectrum as a function
of time for two different values of $\delta f$. Time is scaled by
the director relaxation time $\tau_d$. (a) $\delta f = 0.008\ {\rm
Hz}$ (b) $\delta f = -0.23\ {\rm Hz}$. In both cases, the open
symbols are for the power spectrum computed on the image with the
stripes, with $\sigma_x$ given by the squares and $\sigma_y$ given
by the circles. The closed symbols are for the power spectrum of
the image of the angle map, with $w_x$ given by the squares and
$w_y$ given by the circles.}
\end{figure}

Figure~\ref{sigma} shows the evolution of the width of the power
spectrum for both the pattern ($\sigma_x$ and $\sigma_y$) and the
angle map ($w_x$ and $w_y$). There are a number of features that
become apparent in these plots. First, if one were to assume that
a scaling regime is reached, the growth exponent measured from
$S(k)$ is very different from that measured for $S(q)$. Based on
$\sigma_x$ and $\sigma_y$, one observes very slow growth for the
system. For late times, assuming $\sigma \sim t^n$, the exponents
$n$ are on the order of $0.1$. In contrast, the length scales
characterized by the inverse of $w_x$ and $w_y$ grow faster. In
this case, one obtains growth exponents of $n = 0.33$ or larger.
This difference is consistent with earlier results comparing
$S(k)$, which includes information about the variation in local
wavenumber, to more direct measures of the domain size
\cite{CM95,PD01}.

The second feature of the growth is the differences between the
growth in the x- and y-directions. For the quench with $\delta f =
0.008\ {\rm Hz}$, both $w_x$ and $w_y$ have the same magnitude and
time dependence. This suggests that the system is evolving in a
relatively isotropic manner. For $\sigma_x$ and $\sigma_y$, the
time evolution is similar, though the magnitudes are different.
This suggests that there is more anisotropy in the local
wavenumber distributions than there is in the domains. This
anisotropy contributes to the difference in magnitude between
$\sigma_x$ and $\sigma_y$, but does not impact the time evolution.
For the case of $\delta f = -0.23\ {\rm Hz}$, the situation is
roughly reversed. In this case, $\sigma_x$ and $\sigma_y$ have
roughly the same magnitude throughout the evolution. In contrast,
$w_x$ and $w_y$ have different magnitudes and slightly different
time evolution, at late times. This qualitative difference is
probably due to the increased area of the pattern that is given by
a superposition of zig and zag rolls, and thus, the change in the
nature of the domain boundaries.

\begin{figure}
\includegraphics[width=3.0in]{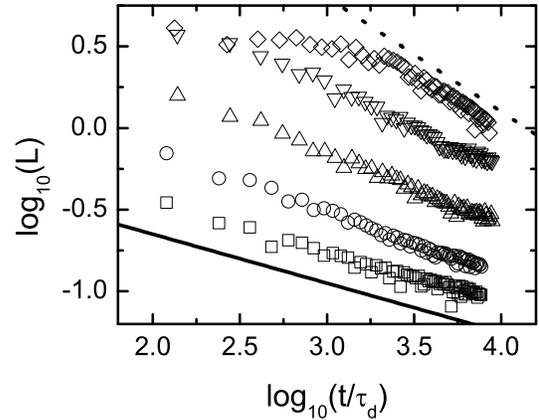}
\caption{\label{length} Total domain wall length as a function of
time for all the quenches studied. Time is scaled by the director
relaxation time, $\tau_d$. The symbols are as follows: $(\square)\
\delta f = 0.008\ {\rm Hz}$, $(\circ)\ \delta f = -0.012\ {\rm
Hz}$, $(\vartriangle)\ \delta f = -0.022\ {\rm Hz}$,
$(\triangledown)\ \delta f = -0.072\ {\rm Hz}$, and $(\lozenge)\
\delta f = -0.23\ {\rm Hz}$. The lines are provided as a guide to
the eye, with the dashed line having a slope of $-0.7$ and the
solid line having a slope of $-0.3$.}
\end{figure}

As an alternative measure of the domain growth, we also considered
the total domain wall length present in a fixed viewing area. This
measure is an {\it isotropic} measure of the domain growth. Its
time evolution is a good measure of how rapidly the overall size
of a domain changes. However, it does not provide any information
about the relative rate of change in the x- and y-direction.
Figure~\ref{length} shows the evolution of the total domain wall
length for a number of values of $\delta f$. Also shown are two
straight lines that provide a guide to the eye. The dashed line
has a slope of $-0.7$, and the solid line has a slope of $-0.3$.
It is clear from this plot that the evolution of the domain size
occurs more rapidly for the larger values of $|\delta f|$.

At this point, it is important to emphasize one of the challenges
of interpreting data used to characterize domain growth. In
studies of domain coarsening, it is standard to assume that a
late-time scaling regime exists in which the lengths
characterizing the domain size grow as a power-law in time. One
experimental difficulty with this assumption is determining when
such a scaling regime is reached. This is especially true in the
studies reported on here, where the average wavenumber is still
evolving in time for all of the conditions that we studied (even
if it is only logarithmic in time). However, independent of
whether or not a true scaling regime has been reached, for similar
times during the evolution of the domains, the data is consistent
with a power-law of the form $l(t) \sim t^n$, where $l$ is some
characteristic length. Therefore, the values of $n$ provide
information about the relative rate of growth of domains under
different circumstances. This is our main motivation for looking
at $n$. Any claims of a true scaling regime would be premature at
this point, though the consistency of the data with power-law
behavior suggests that it is possible such a regime has been
reached.

\begin{figure}
\includegraphics[width=3.0in]{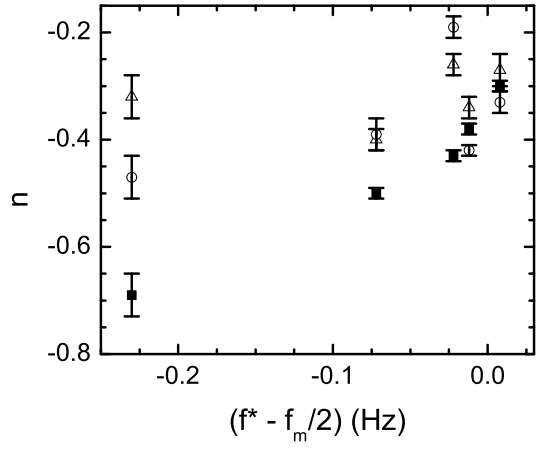}
\caption{\label{exponent} Summary of measured scaling exponents
($n$) for both the domain wall length (solid squares),  $w_x$
(open triangles), and $w_y$ (open circles).}
\end{figure}

Figure~\ref{exponent} summarizes the impact of variations in the
wavenumber on the domain growth by comparing growth exponents $n$
for $w_x$, $w_y$, and the total domain wall length per viewing
area $L$. The results from $\sigma_x$ and $\sigma_y$ exhibited no
strong dependence on $\delta f$. One observes that the growth of
the domains is significantly faster for larger $|\delta f|$, as
measured by the total domain wall length in a given area. In this
case, $n$ increases in magnitude monotonically with $|\delta f|$.
However, the trends in $w_x$ and $w_y$ are less obvious. If
anything, one observes a small change in the magnitude of $n$ for
$w_x$ and $w_y$ as $|\delta f|$ is increased. Also, the growth
appears to become more anisotropic as $|\delta f|$ increases.

\section{Summary}

The experiments reported on here demonstrate the close connection
between the dynamics of domain growth in patterned systems and the
evolution of the background wavenumber. We performed a series of
quenches in which the steady-state average wavenumber after the
quench is varied. The resulting domain growth depends on the
evolution and steady-state value of $<k>$. However, different
measures of the domain growth are impacted to different degrees. A
direct measure of the typical domain size, the average domain wall
length in a fixed viewing area, showed the strongest dependence on
the average wavenumber, with the rate of growth increasing as the
wavenumber is varied away from its ``natural'' value.

If one assumes that the average domain wall length scales as a
power in time, exponents in the range of $0.3$ to $0.7$ are
observed. As discussed in the introduction, such a large variation
in growth exponent is consistent with simulations of the two
different Swift-Hohenberg equations \cite{CM95}. In
Ref.~\cite{CM95}, the calculated exponent is found to increase
from $0.25$ to $0.5$ with a variation in wavenumber of
approximately 20\%. This difference in average wavenumber was
suggested as the source of the different exponents. The work in
Ref.~\cite{CM95} does not directly apply to this system because it
considered the isotropic Swift-Hohenberg equation, and our system
is anisotropic. However, the similarity between the two results
suggests that it is worth further work to determine how universal
the impact of wavenumber variation is on domain growth in
patterned systems.

An interesting feature of the growth is that as one forced the
average wavenumber to be different from the ``natural'' wavenumber
of the system, the rate of growth was increased. In our case, the
maximum variation in wavenumber was only 6\%, but the exponent
changed from $0.3$ to $0.7$. If these results carry over to other
patterned systems, such as diblock copolymers, one can have a
relatively significant impact on the rate of growth of domains by
inducing a change in the average wavenumber away from the
``natural'' wavenumber. This increase in the rate of domain
coarsening could have a practical impact in various processing
situations.

\begin{acknowledgments}

This work was supported by NSF grant DMR-9975479. M. Dennin also
thanks the Research Corporation and Alfred P. Sloan Foundation for
additional funding for this work.

\end{acknowledgments}

\clearpage

\end{document}